\documentclass{elsart}
\usepackage{graphicx}
\usepackage{amsmath}
\usepackage{amssymb}

\begin{document}

\begin{frontmatter}

\title{
In$_{0.75}$Ga$_{0.25}$As on GaAs submicron rings and their
application for coherent nanoelectronic devices }

\author[address1]{Franco Carillo\thanksref{thank1}},
\author[address2]{Giorgio Biasiol},
\author[address1]{Diego Frustaglia},
\author[address1]{Francesco Giazotto},
\author[address3]{Lucia Sorba},
and
\author[address1]{Fabio Beltram}

\address[address1]{NEST-INFM and Scuola Normale Superiore, I-56126 Pisa, Italy}

\address[address2]{NEST-INFM and Laboratorio Nazionale TASC-INFM-CNR, Area Science Park, I-34012 Trieste, Italy}

\address[address3]{NEST-INFM and Laboratorio Nazionale TASC-INFM-CNR, AREA Science Park, I-34012 Trieste, Italy and Universit\`{a} di Modena e Reggio Emilia, I-41100 Modena, Italy}
\thanks[thank1]{
Corresponding author. E-mail: franco.carillo@sns.it}

\begin{abstract}
Electron-phase modulation in magnetic and electric fields will be presented in
In$_{0.75}$Ga$_{0.25}$As Aharonov-Bohm (AB) rings. The zero Schottky
barrier of this material made it possible to nanofabricate
devices with radii down to below $200$~nm without carrier depletion.
We shall present a fabrication scheme based on wet and dry etching
that yielded excellent reproducibility, very high contrast of the
oscillations and good electrical gating. The operation of these
structures is compatible with closed-cycle refrigeration and
suggests that this process can yield coherent electronic circuits
that do not require cryogenic liquids. The InGaAs/AlInAs
heterostructure was grown by MBE on a GaAs substrate \cite{tasc}, and
in light of the large effective g-factor and the absence of the Schottky
barrier is a material system of interest for the investigation of
spin-related effects \cite{desr1,desr2,spint} and the realization of
hybrid superconductor/semiconductor devices \cite{sns}.
\end{abstract}

\end{frontmatter}

\section{Introduction}
In the framework of conventional electronics, the requirement of an increasing
 number of devices per unit area has been the leading motivation
for progressive device shrinking. As far as coherent electronics is concerned
 device shrinking is desirable in order to enhance
performance and to rise the maximum working temperature. Established
technologies for defining nanostructures in two-dimensional electron
gases (2DEGs) are: nano gates, atomic force microscopy lithography
(AFML) \cite{fuhr} and etched side-walls. The first two techniques
where successfully employed in GaAs/AlGaAs-based nanostructures
\cite{ferry}. Etched nanostructures, despite the strong confinement,
have charge depletion layers at the border, the extension of which
depends on the type of semiconductor. In the case of GaAs-based 2DEG
depletion severely limits the smallest size of a device, being in
this case larger than those obtainable using AFML. We want to
demonstrate that the use of InGaAs based 2DEGs overcome this limit
allowing to pattern nanostructures defined by etched side walls down
to $80$ nm line-width providing at the same time a good electric
gating.
\begin{figure}[t]
\begin{center}\leavevmode
\includegraphics[width=1.0\linewidth]{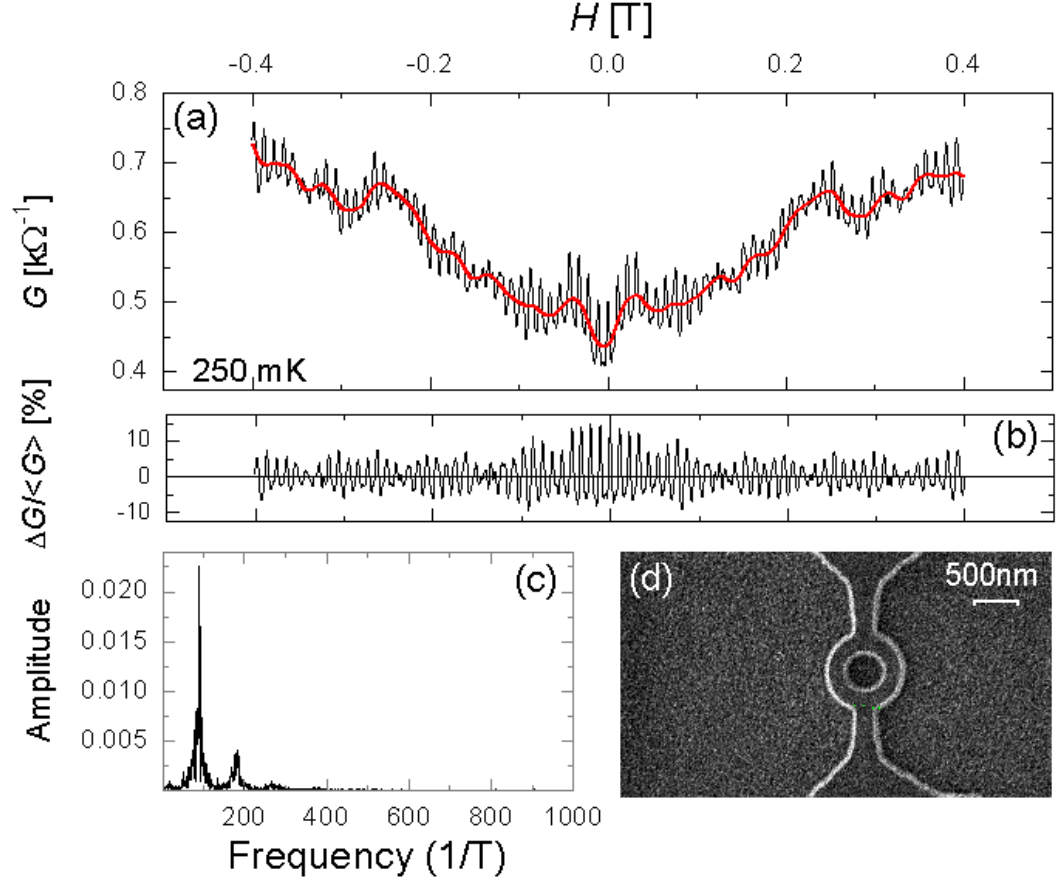}
\caption{(a) AB magnetic field conductance oscillation for a 350-nm-mean-radius
ring shown in panel (d). (b) $G(H)$ curve of panel
(a) where the slowly-varying part (thick line) was subtracted. The
peak to peak conductance oscillation amplitude is around the $20\%$
of the average conductance value.}
\label{nogate}\end{center}\end{figure}

\section{Device description and fabrication process}
The heterostructure employed in this work was grown by MBE on a GaAs
$(100)$-oriented substrate \cite{tasc}. A sequence of
In$_{x}$Al$_{1-x}$As layers of increasing In content was first
deposited in order to ensure lattice matching with the upstanding 10
nm layer of In$_{0.75}$Ga$_{0.25}$As forming the quantum well. A 10-nm-thick
In$_{0.75}$Al$_{0.25}$As spacer separates the well from a
Si -doping layer $(N_\text{{Si}}=1.3\times 10^{12}$ cm$^{-2}$)
placed above. At 4.2 K the electron mobility is $1.1\times 10^5$
cm$^{2}$V$^{-1}$s$^{-1}$, and the carrier density $9.4\times 10^{11}$
cm$^{-2}$, as estimated from Shubnikov-de Haas measurements on a
$60$-$\mu$m-wide Hall bar patterned on the same heterostructure of
the devices presented here. From the mobility and density we deduced
the momentum relaxation length $l_m=1.76\,\mu$m. The ballistic
thermal length is $l_T=\hbar v_F/\pi k_BT=0.46\,\mu$m at $4.2$ K,
where $v_F$ is the Fermi velocity and $T$ is the temperature. The
first fabrication step consisted in the deposition and annealing of
Ni/AuGe/Ni/Au contacts. A Ti mask was then defined on the substrate
surface by electron beam lithography, thermal evaporation of Ti, and
lift-off. A reactive ion etching process was performed to
transfer the geometry of the Ti mask on the semiconductor. The gas
mixture employed for the dry etching was Ar/CH$_4$/H$_2$. The Ti
mask was finally removed by wet chemical etching. Tens of devices
were fabricated on the same sample showing highly-reproducible
structural and transport properties. This was extremely useful
towards the optimization of devices shape, and, in principle, could
be used for tuning circuits parameters.

\section{Coherent transport in In$_{0.75}$Ga$_{0.25}$As-based nanostructures}
Two-probe differential conductance measurements were performed with
standard lock-in technique at $17.4$ Hz. While measuring at $4.2$~K
or higher temperatures, the bias amplitude was kept at $100\,\mu$V,
while at $250$ mK we used $20\,\mu$V. Figure~\ref{nogate} shows the
results for a $350$ nm average radius ring. The conductance $G$
exhibits AB modulations at $250$ mK with a contrast $\Delta
G/G=20\%$, a remarkably high value for a structure defined using dry
etching. This fact joined with the visibility of higher harmonics in
$G(H)$'s Fast Fourier Transform (FFT) (see Fig. \ref{nogate}(c)), indicates that our process
has a small impact on the structural and transport properties of the
patterned nanostructures. The sample electric stability at 250 mK was
tested on a time scale of the order of $15$ hours. The two-probe
conductance $G(B)$  is expected to be symmetric with respect to the
magnetic field. The high symmetry observed in magnetic field further
confirms a good stability of the device. From the frequency value of
the first harmonic peak we deduced an effective radius consistent
with the lithographically-defined average radius. All the rings
tested confirmed the same behavior.

\begin{figure}[t]
\begin{center}\leavevmode
\includegraphics[width=1.0\linewidth]{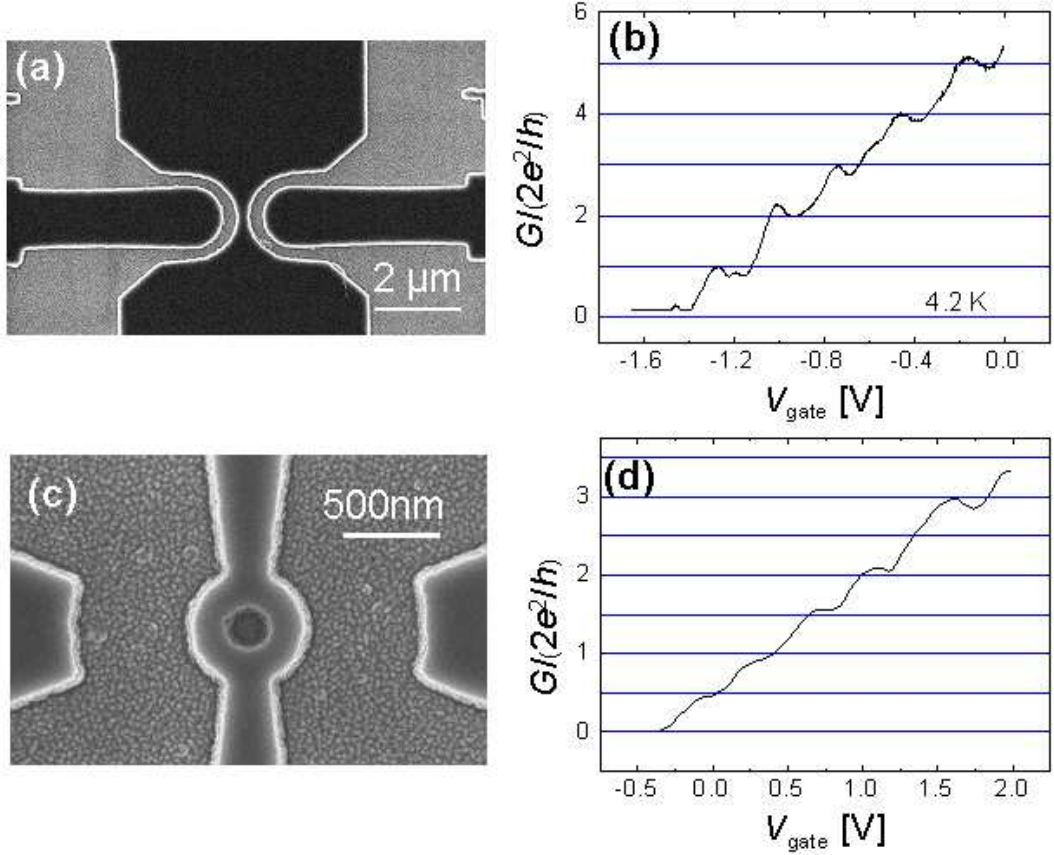}
\caption{(a) SEM micrograph of a $750$-nm-radius and $200$-nm-width
QPC. (b) Conductance quantization as a function of voltage applied
at both side gates of the QPC shown in (a). (c) SEM image of a ring
featuring   $200$ nm mean radius and 120 nm arm width. Side gates
are defined at a distance of $550$ nm apart from each branch. (d)
Zero-bias differential conductance at $4.2$ K for the ring shown in
(c).} \label{CQuant}\end{center}\end{figure}

The strong and sharp lateral confinement provided by the etched
side-walls and the reduced arms line-width lead to a large
separation between energy sub-bands. This, joined with reduced
electronic paths, allowed us to prove coherent transport in a
200-nm-radius ring similar to that of Fig.~\ref{CQuant}(c) up to
$T=10$~K.

\section{Electric gating on In$_{0.75}$Ga$_{0.25}$As}
The absence of the Schottky barrier makes it difficult to realize
electrostatic gating on In$_{0.75}$Ga$_{0.25}$As-based
nanostructures. We defined in-plane side gates patterned onto the
2DES itself. In Fig.~\ref{CQuant}(a) is shown a quantum point
contact (QPC). Five conductance steps are clearly visible at 4.2
K (see Fig.~\ref{CQuant}(b)). The pinch-off was
induced by applying -1.6 V to the gates. At $V=0$ five transverse
 modes are present in the QPC.

Figure~\ref{CQuant}(c) displays the SEM micrograph of a typical
200-nm-radius ring,  while Fig. ~\ref{CQuant}(d) shows its
differential conductance vs gate voltage. By applying a voltage
sweep at both side gates the conductance exhibits well-defined steps
corresponding to multiples of $e^2/h$ (Fig.~\ref{CQuant}(d)). This
can be explained considering two series quantum point contact at the
entrance and at the exit of the ring. The same behavior have been
observed also in GaAs/AlGaAs-based rings \cite{Pedersen}.

\begin{figure}[t]
\begin{center}\leavevmode
\includegraphics[width=1.0\linewidth]{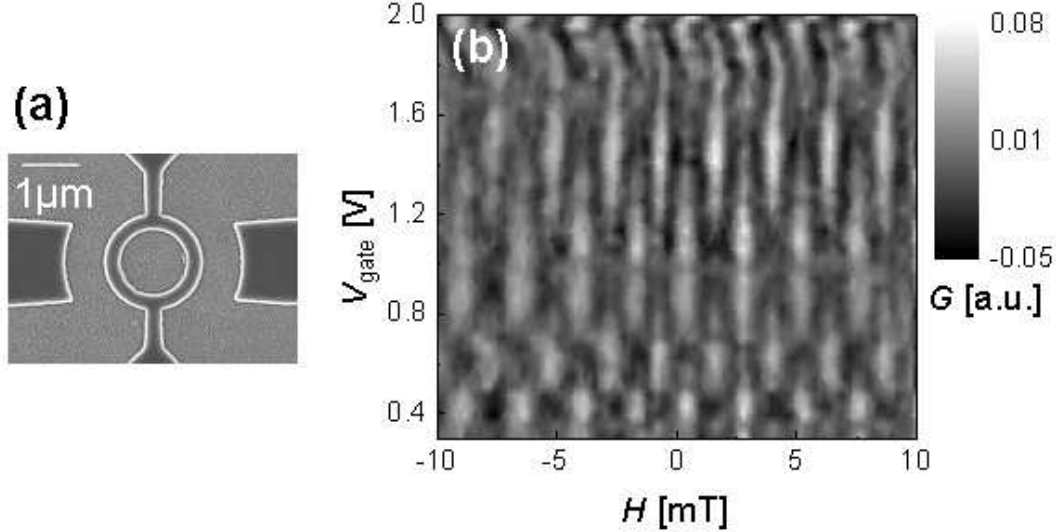}
\caption{(a) SEM micrograph of a 750-nm-radius ring featuring arm
width of 200 nm. (b) Density plot of the  conductance  vs magnetic
field (\emph{x} axis) and voltage applied at one of the two gates
(\emph{y} axis). The other gate was biased at 1 V. The ac bias was
set at $20\,\mu$ V and temperature at $280$ mK.}
\label{interferBig}\end{center}\end{figure}

\section{Voltage-controlled interferometers}
\begin{figure}[t]
\begin{center}\leavevmode
\includegraphics[width=1.0\linewidth]{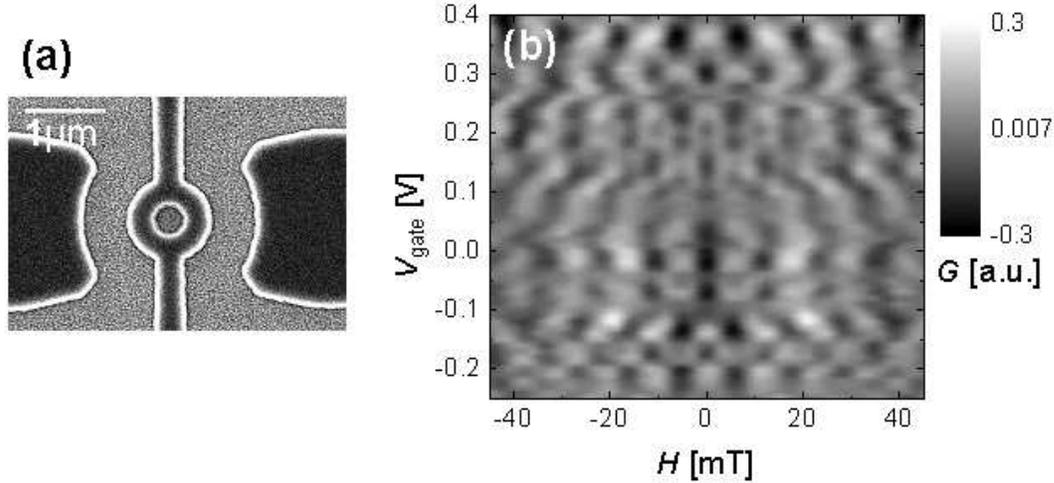}
\caption{(a) SEM image of a 350-nm-radius ring featuring arm width
of 200 nm. (b) Density plot of the ring conductance like in
Fig.~\ref{interferBig}(b).}\label{interferSm}\end{center}\end{figure}

Many  theoretical \cite{sols,cahay,takai} and experimental
\cite{schaptrans} works focused on the possibility to control
devices conductance acting electrostatically on the quantum phase of
charge carriers. In such systems it is not necessary to modify  the
charge density itself, like in conventional field-effect
transistors, and this should result in higher switching speed and
extended frequency range. Furthermore, phase-controlled devices
could be fruitfully employed in complex coherent circuits, where
phase manipulation is exploited for complex operations rather than
the more simple control of conductance. One relevant geometry that
allows the implementation of electrostatically-controlled
interferometers is represented by a 2DEG ring where one of the two
arms is capacitively-coupled to a nearby gate \cite{El-AB,shapers}.
Using the previously described process we fabricated rings of
different sizes with gates like those shown in Figs.~\ref{CQuant} to
\ref{interferSm}. By applying a bias to one of the gates while
keeping the other at a fixed voltage it is possible to change the
phase acquired by an electron traversing the arm closer to the
former gate \cite{cahay,takai}. The effect of this extra-phase can
be detected by analyzing the conductance vs magnetic field and gate
voltage \cite{shapers}. Figure~\ref{interferBig} shows the SEM
micrograph of a $750$ nm ring and the density plot of its
conductance at $280$ mK as a function of the voltage applied at one
side gate (vertical axis) and of the external magnetic field
(horizontal axis). By changing the voltage at the side gate, the
sequence of maxima and minima in magnetic field (bright and dark
regions, respectively) are shifted by a half period $h/2e$. This
shift occurs several times in the considered voltage range.
Figure~\ref{interferSm} shows the conductance density plot of a
$350$-nm-radius ring. In this case, due to the different gates
geometry and to the reduced device size, the bias influences both
the leads and the farthest arm of the ring, resulting in an
irregular and distorted pattern. Density plots of
Fig.~\ref{interferBig} and Fig.~\ref{interferSm} were obtained by
subtracting the slowly-varying part of the conductance in order to
evidence the oscillations in magnetic and electric field.

\section{Conclusion}
We demonstrated a highly-reproducible process based on
In$_{0.75}$Ga$_{0.25}$As 2DES, compliant with large scale
integration and capable to reliably produce rings with average
radius as small as $200$ nm. The electric gating implemented with such
a process efficiently controls charge density as well as induces
conductance quantization. Voltage-controlled electron
interferometers were demonstrated for different device size and
geometries.

\end{document}